\documentclass[twocolumn,showpacs,preprintnumbers,amsmath,amssymb]{revtex4}

\usepackage[dvips]{graphicx}
\usepackage{epsfig}
\usepackage{dcolumn}
\usepackage{bm}

\newcommand{\nn}{\nonumber}

\def\sh{\hat{s}}
\def\th{\hat{t}}
\def\uh{\hat{u}}

\def\gev{\,{\rm GeV}}

\begin{document}

\preprint{WU B 02-08}
\preprint{PITHA 02/18}
\preprint{hep-ph/0212138}

\title{Proton mass effects in 
wide-angle Compton scattering}

\author{M.\ Diehl, Th.\ Feldmann}
 \affiliation{Institut f\"ur Theoretische Physik E, RWTH Aachen,
    52056 Aachen, Germany}

\author{H.W.\ Huang}%
\affiliation{Department of Physics, University of Colorado, 
   Boulder, CO 80309-0390, USA}

\author{P.\ Kroll}
\affiliation{ Fachbereich Physik, Universit\"at Wuppertal, 42097
Wuppertal, Germany}

\date{December 2002}

\begin{abstract}
We investigate proton mass effects in the handbag approach
to wide-angle Compton scattering. We find that
theoretical uncertainties due to the proton 
mass are significant for photon energies 
presently studied at Jefferson Lab. With the proposed energy upgrade 
such uncertainties will be clearly reduced.  
\end{abstract}

\pacs{13.60.Fz}

\maketitle

In Refs.~\cite{DFJK1,DFJK2,hkm} we have investigated the handbag
approach to wide-angle Compton scattering off protons, $\gamma p\to
\gamma p$.  Analogous results have been obtained in Ref.~\cite{rad}.
In the handbag approach the Compton amplitude is given by a hard
scattering $\gamma q\to \gamma q$ at parton level multiplied by soft
Compton form factors describing the emission and reabsorption of the
quark by the proton.  The kinematical requirement for the
applicability of this approach is that the Mandelstam variables $s$,
$-t$ and $-u$ are large compared to a typical hadronic scale of order
$\Lambda^2=1 \gev^2$. This implies $s, -t, -u \gg m^2$, where $m$ is
the proton mass. At Jefferson Lab (JLAB) there are ongoing experiments
to measure the Compton cross section and certain spin transfer
parameters \cite{Chen:az}.  Presently available beam energies are
however not very high.  In this note we investigate, as an example,
the role of a non-negligible target mass in the handbag approach.  An
important issue in this context is the way to relate the dynamical
variables of the approach to the external kinematics, determined by
the experimental conditions.  This relation is not unambiguous, which
is one of the sources of theoretical uncertainties in the handbag
approach.  We study three different approximations and
take the differences in their predictions as a measure of the
theoretical uncertainty, which should be taken into account in
attempts to extract the Compton form factors from experimental
data.  

The external kinematics is determined by the beam energy
$E_{L}^\gamma$ in the laboratory and by the scattering angle $\theta$
in the center of mass frame.  These quantities fix the external
Mandelstam variables by
\begin{eqnarray}
s&=& 2 m E_L^\gamma + m^2 \, , \nn\\
t&=&-\frac{s}{2} (1-\cos{\theta})\, (1-m^2/s)^2 \, ,  \nn\\
u&=& 2 m^2 -s -t \, .
\label{mandelstam}
\end{eqnarray} 
These variables should not be changed or approximated in a theoretical
calculation.  Keeping this in mind we suggest a separate treatment of
the kinematical factors from phase space and flux and of the
scattering amplitude, which contains the dynamics.  In the handbag
approach the Compton cross section then reads \cite{DFJK1,hkm}
\begin{eqnarray}
\lefteqn{\frac{d\sigma}{dt}
= \frac{\pi \alpha_{\rm elm}^2}{(s-m^2)^2}\;
}
\\[0.25em]
&& {}\times \left[ \frac{(\sh - \uh)^2}{|\sh \uh|} \Big( R_V^2(\th)
                        - \frac{\th}{4m^2} R_T^2(\th) \Big)
                 + \frac{(\sh + \uh)^2}{|\sh \uh|} R_A^2(\th) \right]
\nn
\label{pred}
\end{eqnarray}
to leading $\mathcal{O}(\alpha_s)$, where $R_V$, $R_A$ and $R_T$ are
the Compton form factors.  The one-loop corrections to the hard
scattering have been evaluated in \cite{hkm}. They were found to be
small in the backward hemisphere and increased up to about 30\% for
$\cos\theta = 0.6$.  The Mandelstam variables $\sh$, $\th$, $\uh$
refer to the partonic subprocess $\gamma q\to \gamma q$.  They
coincide with the external variables $s$, $t$, $u$ up to corrections
of order $\Lambda^2 /s$.  To calculate these consistently is beyond
the accuracy of the approach in its present form.  In particular,
different choices for $\sh$, $\th$, $\uh$ lead to different results
for the cross section at finite $s$.

We investigate the numerical effects of this ambiguity in three
different scenarios.  For the beam energy we take
$E_L^\gamma=4.3\gev$, corresponding to $s=8.97 \gev^2$, where there
will soon be data from JLAB.  We compare this with the situation for
the energy $E_L^\gamma=12\gev$ of the proposed JLAB upgrade.  We take
the form factors $R_V$ and $R_A$ modeled in \cite{DFJK1} using the
overlap of light-cone wave functions and its connection to parton
distributions and elastic form factors.  From the overlap
representation one expects a similar suppression of $R_T/R_V$ as for
the ratio $F_2/ F_1$ of electromagnetic Pauli and Dirac form factors.
For simplicity, we neglect $R_T$ when evaluating the Compton cross section.

\begin{description} 
\item[Scenario 1:]
\begin{equation}
\sh = s\,, \qquad \th=t\,, \qquad  \uh=u\,.  
\label{scen1}
\end{equation}
The subprocess amplitude leading to (\ref{pred}) was calculated in the
approximation of massless on-shell quarks, so that the sum of internal
Mandelstam variables is zero.  In this scenario this only holds
approximately, since $\hat{s}+\hat{u}+\hat{t}=2m^2$.
\item[Scenario 2:]
\begin{equation}
\hat{s}=s-m^2\,, \qquad \hat{t}=t\,, \qquad \hat{u}=u-m^2\,,
\end{equation}
where now $\hat{s}+\hat{u}+\hat{t}=0$.
\item[Scenario 3:]
\begin{equation}
\sh = 2m\, E_L^\gamma\,, \qquad 
\th = -\frac{\sh}{2}\, (1-\cos{\theta})\,, \qquad
\uh=-\sh -\th\,.
\label{scen3}
\end{equation}
Notice that in this case one has $\th \neq t$.
\end{description}

\begin{figure*}
\centerline{\includegraphics[width=0.9\textwidth,bb= 50 210 510 390]{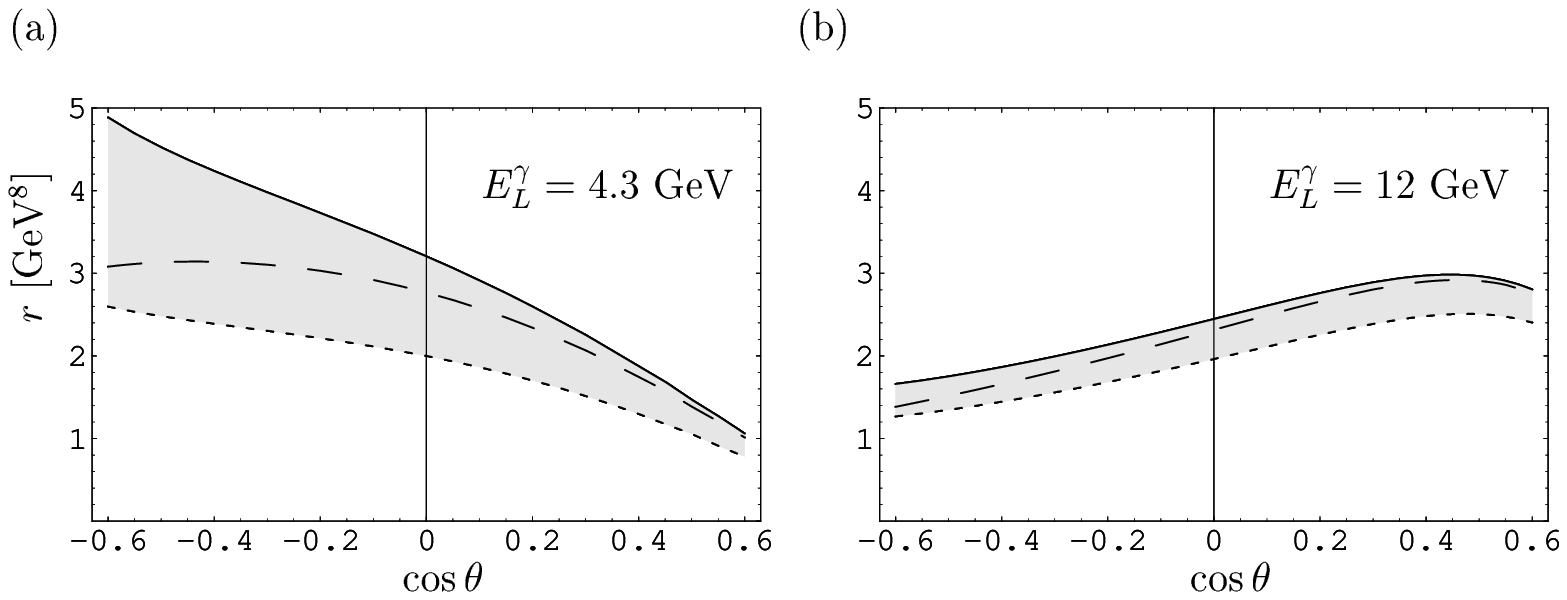}}
\caption{\label{fig:cross} (a) The ratio $r$ defined in
Eq.~(\ref{ratio-def}) at $E_L^\gamma=4.3\gev$ for scenarios~1~(full),
2~(long-dashed), 3~(short-dashed).  The form factors $R_V$ and $R_A$
are taken from the model in \protect\cite{DFJK1} and $R_T$ is
neglected.  (b) The same for $E_L^\gamma=12\gev$.}
\end{figure*}

Numerical results for the Compton cross section evaluated from
(\ref{pred}) without the $R_T$ and $\alpha_s$ corrections are shown in
Fig.\ \ref{fig:cross}(a) for the three scenarios.  We plot the ratio
\begin{equation}
\label{ratio-def}
r = t^4 \frac{d\sigma/dt}{d\sigma_{\mathrm{KN}}/dt} ,
\end{equation}
where
\begin{eqnarray}
&& \frac{d\sigma_{\mathrm{KN}}}{dt} \ = \
\frac{2\pi \alpha_{\rm elm}^2}{(s-m^2)^2}\;
\\[0.2em]
&& \times
\left[ 
  - \frac{s-m^2}{u-m^2} - \frac{u-m^2}{s-m^2} 
  + \frac{4 m^2 t\, (m^4 - su)}{(s-m^2)^2 (u-m^2)^2}
\right]
\nn
\end{eqnarray}
is the Klein-Nishina cross section for a point-like proton.  The ratio
$r$ essentially measures an average of $(t^2 R_V)^2$ and $(t^2
R_A)^2$.  These scaled form factors are expected to depend only weakly
on $t$ in the kinematical range considered here \cite{DFJK1,DFJK2}.

For a beam energy of $4.3 \gev$ the differences between the 
cross sections evaluated in the three scenarios are
moderate at small scattering angles but grows up to a factor 2 for
backward angles.  With a beam energy of $12 \gev$ instead, the
ambiguities are small for all angles considered, as shown in
Fig.~\ref{fig:cross}(b).

In \cite{DFJK1,DFJK2,hkm} we plotted the scaled Compton cross section
$s^6 d\sigma/dt$ with the squared proton mass neglected in both the
internal and \emph{external} variables, i.e.\ we used scenario 3
together with $s=\sh$, $t=\th$, $u=\uh$ and replaced $(s-m^2)^2$ with
$s^2$ in (\ref{pred}).  In this case the scattering angle $\theta$ and
the scaling parameter $s^6$ multiplying the differential cross section
do not correspond to the experimentally measured quantities. As a
consequence both data and theoretical predictions in the plots of
\cite{DFJK1,DFJK2,hkm} were shifted. In other words we plotted $(2 m
E_L^\gamma)^6 d\sigma/dt$ against $1 + t /(m E_L^\gamma)$ rather than
$s^6 d\sigma/dt$ against $\cos\theta$.  We realize that this is a
rather confusing procedure which should be avoided in future
presentations (we thank Travis Brooks and Lance Dixon for drawing our
attention to this problem). Nevertheless, if our kinematical
requirements of $s, -t, -u \gg \Lambda^2$ are well satisfied the
expressions given in \cite{DFJK1,DFJK2,hkm} are correct. The
subtleties of internal vs.\ external Mandelstam variables matter only
for energies as low as those currently available at JLAB, where the
application of the handbag approach requires some care and is perhaps
more qualitative.

\begin{figure*}
\centerline{\includegraphics[width=0.9\textwidth,bb = 50 210 510 390]{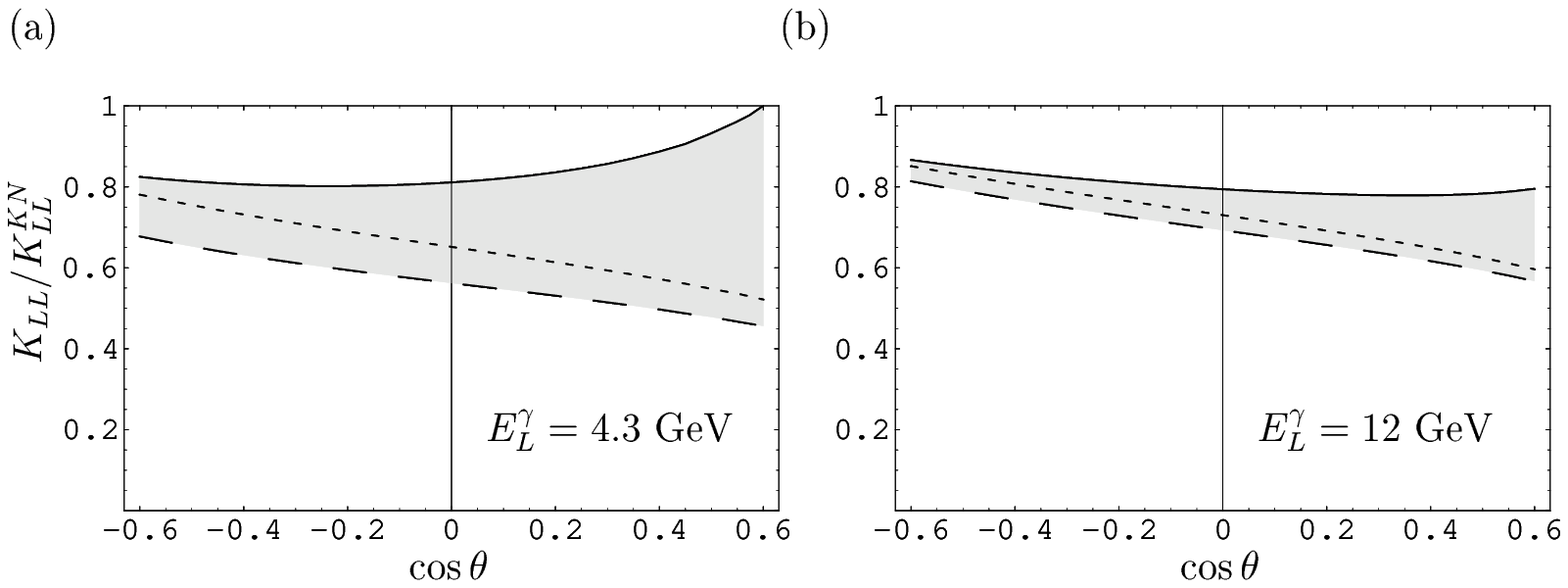}}
\caption{\label{fig:all} (a) The ratio $K_{LL} /K_{LL}^{\mathrm{KN}}$
of helicity correlations at $E_L^\gamma=4.3\gev$ for scenarios~1
(full), 2 (long-dashed), 3 (short-dashed). $R_V$ and $R_A$ are taken
from the model in \protect\cite{DFJK1} and $R_T$ is neglected.  (b)
The same for $E_L^\gamma=12\gev$.}
\end{figure*}

Another interesting quantity is the correlation between the helicities
of the incoming photon and the incoming ($A_{LL}$) or the outgoing
($K_{LL}$) proton in the c.m. In the handbag approach these
parameters are given to leading $\mathcal{O}(\alpha_s)$ by
\cite{DFJK2,hkm} 
\begin{eqnarray}
  \label{eq:all}
\lefteqn{
\frac{d\sigma}{dt} A_{LL} = \frac{d\sigma}{dt} K_{LL} =
\frac{2\pi \alpha_{\rm elm}^2}{(s-m^2)^2} 
}
\nn \\[0.3em]
&\times&
\frac{\sh^2 - \uh^2}{|\sh\uh|}\, R_A(\th)\, \Big( R_V(\th)
               + \frac{\th}{\sh + \sqrt{-\sh\uh}}\, R_T(\th) \Big) .
\end{eqnarray}
The corresponding expressions for a point-like proton are
\begin{eqnarray}
&& 
\frac{d\sigma^{\mathrm{KN}}}{dt} \, A_{LL}^{\mathrm{KN}} =
\frac{2\pi \alpha_{\mathrm{em}}^2}{(s-m^2)^2} \,
\\[0.25em]
&& \quad \times
\left[ - \frac{s-m^2}{u-m^2} + \frac{u-m^2}{s-m^2}
       - \frac{2 m^2 t^2  (s-u)}{(s-m^2)^2 \, (u-m^2)^2} \right] \ ,
\nonumber \\[0.5em]
&&
\frac{d\sigma^{\mathrm{KN}}}{dt} \, K_{LL}^{\mathrm{KN}} = 
\frac{2\pi \alpha_{\mathrm{em}}^2}{(s-m^2)^2} \,
\nonumber \\[0.25em]
&& \quad \times
\left[ - \frac{s-m^2}{u-m^2} + \frac{u-m^2}{s-m^2}
       - \frac{4 m^2 t^2  (m^4 -su)}{(s-m^2)^3 \, (u-m^2)^2} \right]
\, ,
\nonumber
\end{eqnarray}
which reduces to $A_{LL}^{\mathrm{KN}} = K_{LL}^{\mathrm{KN}} =
(s^2-u^2)/(s^2+u^2)$ in the massless limit.  In Fig.\ \ref{fig:all} we
show the helicity correlation $K_{LL}$ evaluated for the three
scenarios, normalized to the Klein-Nishina value.  This quantity is
essentially a measure of $R_A/ R_V$, as can be seen from
\begin{equation}
\frac{K_{LL}}{K_{LL}^{\mathrm{KN}}} \simeq \frac{R_A}{R_V} \,
\left[ 1 - \frac{t^2}{2(s^2 + u^2)} \Big( 1 - \frac{R_A^2}{R_V^2}
                                        \Big) \right]^{-1}  ,
\end{equation}
where we have neglected $R_T$ and kinematic corrections of order $m^2
/s$.  Note that the kinematical prefactor in brackets is at most 0.3
for $\cos\theta \ge -0.6$ and $s\gg m^2$.  Fig.\ \ref{fig:all} shows
that for present JLAB energies the uncertainties related to the proton
mass are sizeable, while at $E_L^\gamma=12$~GeV the effect is small.

\begin{figure*}
\centerline{\includegraphics[width=0.9\textwidth, bb= 50 210 510 390]{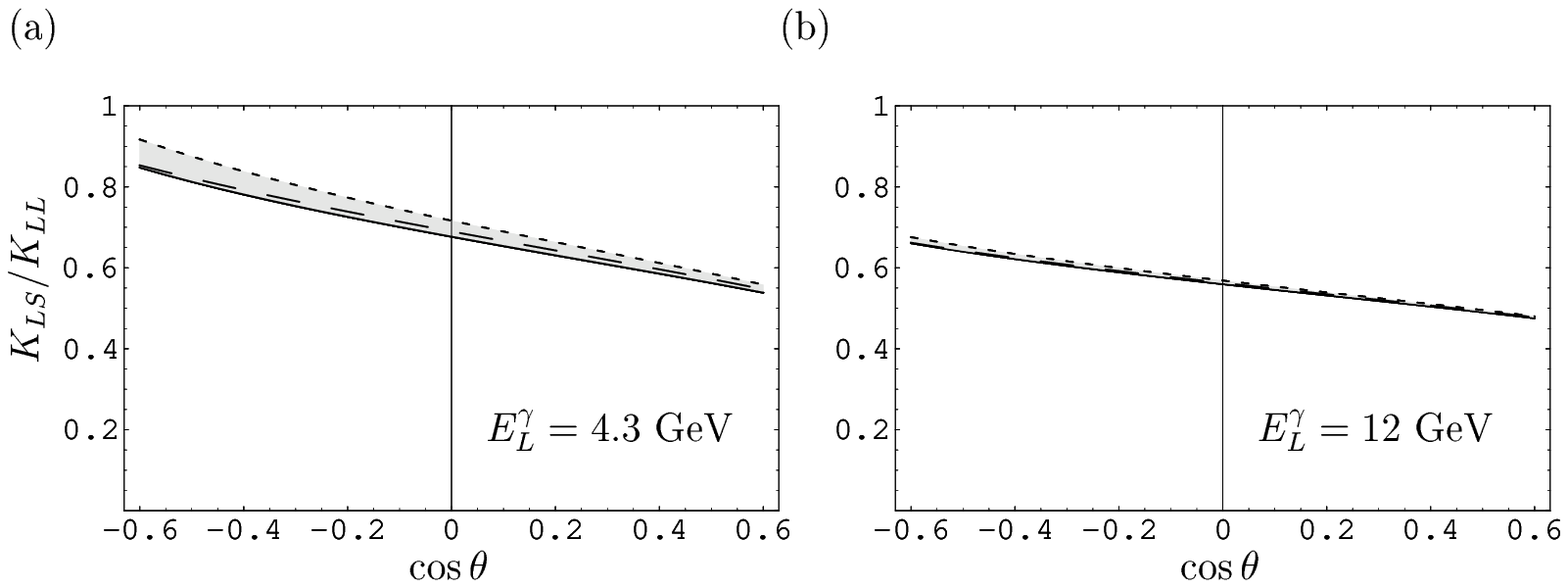}}
\caption{\label{fig:kls} (a) The ratio $K_{LS}/K_{LL}$ of helicity
correlations at $E_L^\gamma=4.3\gev$ for scenarios~1~(full), 2~(long-dashed),
3~(short-dashed).  $R_V$ is taken from the model in
\protect\cite{DFJK1} and $R_T$ is estimated from 
(\protect\ref{rt-ansatz}).  (b) The same for $E_L^\gamma=12\gev$.}
\end{figure*}

JLAB will also measure the correlation $K_{LS}$ between the helicity
of the incoming photon and the sideways polarization of the
outgoing proton. In the handbag approach it is given by \cite{hkm}
\begin{eqnarray}
  \label{eq:kls}
\lefteqn{
\frac{d\sigma}{dt} K_{LS} =
\frac{2\pi \alpha_{\rm elm}^2}{(s-m^2)^2} 
}
\\[0.3em]
&\times&
\frac{\sh^2 - \uh^2}{|\sh\uh|}\, \frac{\sqrt{-\th\,}}{2m}\, 
  \Big( R_T(\th) + \frac{4m^2}{\sh + \sqrt{-\sh\uh}}\, R_V(\th) \Big) \,
  R_A(\th)
\nn 
\end{eqnarray}
to leading $\mathcal{O}(\alpha_s)$, with the sign convention
detailed in \cite{hkm}.  Contrary to the previous
observables, $K_{LS}$ is rather sensitive to the tensor Compton form
factor $R_T$. It is convenient to consider the ratio $K_{LS}/K_{LL}$
where in the handbag approach the form factor $R_A$ drops out.
Introducing the abbreviation
\begin{equation}
  \kappa_T(\th) \equiv \frac{\sqrt{-\th}}{2m} \,
  \frac{R_T(\th)}{R_V(\th)}
\, ,
\end{equation} 
this ratio can be expressed as \cite{hkm}
\begin{equation}
 \frac{K_{LS}}{K_{LL}} 
=
\kappa_T \bigg(1+\frac{2 m \sqrt{-\th}}{\sh + \sqrt{-\sh \uh}} \,
  \frac{1}{\kappa_T} \bigg) 
\bigg(1-\frac{2 m \sqrt{-\th}}{\sh + \sqrt{-\sh \uh}} \,
  {\kappa_T} \bigg)^{-1} \! .
\label{KLSratio}
\end{equation}
A rough estimate for 
the quantity $\kappa_T$ can be obtained by considering
the analogy between the ratio $R_T/ R_V$ and its
electromagnetic counterpart $F_2/F_1$.
One may, for instance, assume that
\begin{equation}
  \label{rt-ansatz}
\kappa_T(\th) 
\simeq
\frac{\sqrt{-\th}}{2m} \, \frac{F_2(\th)}{F_1(\th)} 
\approx 0.37 \, ,
\end{equation}
where the numerical value on the r.h.s.\
is taken from the measurement of $F_2/F_1$ for 
$-t = 1 \div 5.6 \gev^2$ \cite{Gayou:2001qd}.
(Note, however, that on the basis of the previous SLAC measurement
of $F_2/F_1$ \cite{Andivahis:1994rq} one would rather conclude that 
$\kappa_T \propto m/\sqrt{-t}$.
For a detailed discussion of the uncertainties in the
measurement of the Pauli form factor see \cite{Arrington:2002cr}.)
We use (\ref{rt-ansatz}) to estimate $K_{LS}$ in the handbag approach
and plot $K_{LS}/K_{LL}$ for the three scenarios in Fig.~\ref{fig:kls}.  
We observe that the ratio $K_{LS}/K_{LL}$ 
is rather insensitive to the proton mass effects already at
the present JLAB energy (in contrast,
the predictions for $K_{LS}$ alone suffer from the same uncertainties
as for the other Compton observables discussed above).
Measuring the ratio $K_{LS}/K_{LL}$ at present JLAB energies,
and solving for $\kappa_T(\th)$ in (\ref{KLSratio})  
thus enables us to determine the ratio $R_T/R_V$ and to test the 
analogy with $F_2/F_1$ in (\ref{rt-ansatz}).

In conclusion, we found that finite proton mass effects severely limit
the quantitative test of the handbag approach and the extraction of
the Compton form factors in wide angle Compton scattering for present
JLAB energies. An exception is the ratio $K_{LS}/K_{LL}$ which turns
out to be rather insensitive to finite proton mass effects, and 
can be used to determine the ratio of Compton form factors
$R_T/R_V$. Qualitative features, like the sign
and order of magnitude of the helicity correlations $K_{LL}$ or
$K_{LS}$, are not affected either. 
For higher photon energies as projected
for JLAB the theoretical uncertainties from the proton mass become
reasonably small.



\begin{thebibliography}{99}

\bibitem{DFJK1} M.~Diehl, T.~Feldmann, R.~Jakob and P.~Kroll,
Eur.\ Phys.\ J.\ C {\bf 8}, 409 (1999)
[hep-ph/9811253].

\bibitem{DFJK2} M.~Diehl, T.~Feldmann, R.~Jakob and P.~Kroll,
Phys.\ Lett.\ B {\bf 460}, 204 (1999)
[hep-ph/9903268].

\bibitem{hkm} H.~W.~Huang, P.~Kroll and T.~Morii,
Eur.\ Phys.\ J.\ C {\bf 23}, 301 (2002)
[hep-ph/0110208].

\bibitem{rad} A.\ V.\ Radyushkin, 
Phys.\ Rev.\ D {\bf 58}, 114008 (1998)
[hep-ph/9803316].

\bibitem{Chen:az}
J.~P.~Chen {\it et al.}  [Jefferson Lab Hall A Collaboration],
``Exclusive Compton Scattering On The Proton,''
Jefferson Lab preprint PCCF-RI-99-17.

\bibitem{Gayou:2001qd}
O.~Gayou {\it et al.}  [Jefferson Lab Hall A Collaboration],
Phys.\ Rev.\ Lett.\  {\bf 88}, 092301 (2002)
[nucl-ex/0111010].

\bibitem{Andivahis:1994rq}
L.~Andivahis {\it et al.},
Phys.\ Rev.\ D {\bf 50}, 5491 (1994).

\bibitem{Arrington:2002cr}
J.~Arrington,
hep-ph/0209243.

\end{thebibliography}
\end{document}